\documentclass[
superscriptaddress,
preprint,
amsmath,amssymb,
aps,
pra,
]{revtex4-2}

\usepackage{graphicx} 
\usepackage{dcolumn} 
\usepackage{bm} 
\usepackage{hyperref}
\usepackage{adjustbox}
\usepackage{booktabs}
\usepackage[section]{placeins}
\usepackage{siunitx}
\usepackage[showframe=false]{geometry}

\begin{document}

%\preprint{APS/123-QED}
%4500 words
\title{Transverse flow under oscillating stimulation in helical square ducts with cochlea-like geometrical curvature and torsion}

\author{N.C.~Harte}
\affiliation{ARTORG Center for Biomedical Engineering Research, University of Bern, Bern, Switzerland}
\affiliation{Department of Otorhinolaryngology, Head and Neck Surgery, Inselspital, Bern University Hospital, Bern, Switzerland}

\author{D.~Obrist}
\affiliation{ARTORG Center for Biomedical Engineering Research, University of Bern, Bern, Switzerland}

\author{M.~Caversaccio}
\affiliation{ARTORG Center for Biomedical Engineering Research, University of Bern, Bern, Switzerland}
\affiliation{Department of Otorhinolaryngology, Head and Neck Surgery, Inselspital, Bern University Hospital, Bern, Switzerland}

\author{G.P.R.~Lajoinie}
\affiliation{Physics of Fluids Group, Department of Science and Technology, MESA+ Institute for Nanotechnology, Technical Medical (TechMed) Center, University of Twente, Enschede, The Netherlands}

\author{W.~Wimmer}
\email{wilhelm.wimmer@unibe.ch}
\affiliation{ARTORG Center for Biomedical Engineering Research, University of Bern, Bern, Switzerland}
\affiliation{Department of Otorhinolaryngology, Head and Neck Surgery, Inselspital, Bern University Hospital, Bern, Switzerland}
\affiliation{Technical University of Munich, Germany; TUM School of Medicine, Klinikum rechts der Isar, Department of Otorhinolaryngology}

\begin{abstract}
The cochlea is our fluid-filled organ of hearing with a unique spiral shape.
The physiological role of this shape remains unclear.
Previous research has paid only little attention to the occurrence of transverse flow in the cochlea, in particular in relation to the cochlea's shape. 
To better understand its influence on fluid dynamics, this study aims to characterize transverse flow due to harmonically oscillating axial flow in square ducts with curvature and torsion, similar to the shape of human cochleae.
Four geometries were investigated to study curvature and torsion effects on axial and transverse fluid flow components. 
Twelve frequencies from 0.125 Hz to 256 Hz were studied, covering infrasound and low-frequency hearing, with mean inlet velocity amplitudes representing levels expected for normal conversations or louder situations.
Our simulations show that torsion contributes significantly to transverse flow in unsteady conditions, and that its contribution increases with increasing oscillation frequencies. 
Curvature has a small effect on transverse flow, which decreases rapidly for increasing frequencies. 
Strikingly, the combined effect of curvature and torsion on transverse flow is greater than expected from a simple superposition of the two effects, especially when the relative contribution of curvature alone becomes negligible. 
These findings could be relevant to understand physiological processes in the cochlea, including metabolite transport and wall shear stresses.
Further studies are needed to investigate possible implications on cochlear mechanics.
\end{abstract}

\maketitle

\clearpage
\section{Introduction}
The cochlea, our organ of hearing, is a fluid-filled structure with a peculiar spiral shape. 
Despite its importance, the physiological role of its shape in sound transmission from the cochlear fluids to the sensory epithelium remains unclear.
Performing experimental investigations is challenging because of the limited access to the cochlea and its small size.
Therefore, existing literature is based on numerical simulations or theoretical studies. 
For example, researchers have simulated the interaction between the cochlear fluids and the basilar membrane (or the organ of Corti) in realistic geometries \citep{givelberg2003comprehensive, bohnke19993d,cai2005effects}, while others provide theoretical insight based on idealized geometries \citep{viergever1978basilar,Loh1983,Steele1985,Manoussaki2006}.
\citet{Manoussaki2008} showed that the spiral shape redistributes wave energy along the radial direction, altering the radial vibration profile of the various cochlear structures and affecting low-frequency hearing. 
However, geometry-related secondary flow phenomena, especially the occurrence of transverse flow (i.e., flow in the cross-section of the cochlea) have received little attention. 

Since the physiology of the ear is fundamentally coupled to fluid dynamic processes, a more detailed understanding of the influence of geometric properties on the flow field is desirable.
In this context, possible mechanisms could be secondary phenomena introduced by transverse flow.
The combination of oscillating axial flow and transverse flow caused by geometry leads to steady streaming effects and net transport of particles \citep{Riley2008}.
Therefore, cross-sectional mixing and longitudinal streaming caused by transverse flow could be relevant for the transport of metabolites in the cochlea \citep{Song2021}. 
In the context of medical treatment, intracochlear fluid-borne mass transport (drug delivery) is an active area of research \citep{Obrist2019,Sumner2021}. 
Moreover, transverse flow can generate wall shear stresses and pressures that could reach magnitudes that may be considered physiologically relevant in the cochlea.

Curvature and torsion are elementary properties that can be used to describe duct geometries by the centerline. 
Relatively simple shapes can generate complex transverse flow patterns. 
The geometry of a toroidal duct, only exhibiting curvature, causes pairs of rotating flow cells oriented in the transverse plane, known as Dean flow \citep{Dean1928}. 
Inside twisted straight ducts, which only contain torsion, saddle flow patterns are generated depending on the cross-section of the duct \citep{Kheshgi1992,Germano1989}. 
More complex phenomena arise in helical ducts with curvature and torsion combined, as has been extensively studied by \citet{Bolinder1996} and others \citep{Gammack2001,Tuttle1990,Thomson2001} under laminar steady flow conditions.
Transverse flow occurring under oscillatory stimulation in toroidal circular ducts was studied by \citet{lyne1971unsteady} and \citet{sudo1992secondary}.

Motivated by the question of the physiological role of the cochlear shape, the aim of this study is to fundamentally characterize transverse flow phenomena under harmonic oscillation in square ducts with curvature and torsion reflecting the shape of human cochleae.
Computational fluid dynamics (CFD) is used to simulate flow at oscillation frequencies within the infrasonic and low-frequency hearing range of humans because the apical region of the cochlea, characterized by a high degree of curvature and torsion, is particularly sensitive to these frequencies.
The results of this study include axial velocity profiles, transverse flow patterns, relative transverse flow magnitudes and phase differences between pressure and velocity. 
Interestingly, we find that torsion remains the dominant contributor to transverse flow under oscillating stimulation, similar to its role under steady flow conditions. 
Additionally, we observe that the combined contribution of curvature and torsion to transverse flow is larger than expected based on a simple superposition of the two effects.

\section{Methods}
\subsection{Duct Geometries}
\label{ch:DuctGeometries}
\begin{figure*}[htbp!]
\centering
\includegraphics[width=0.9\textwidth]{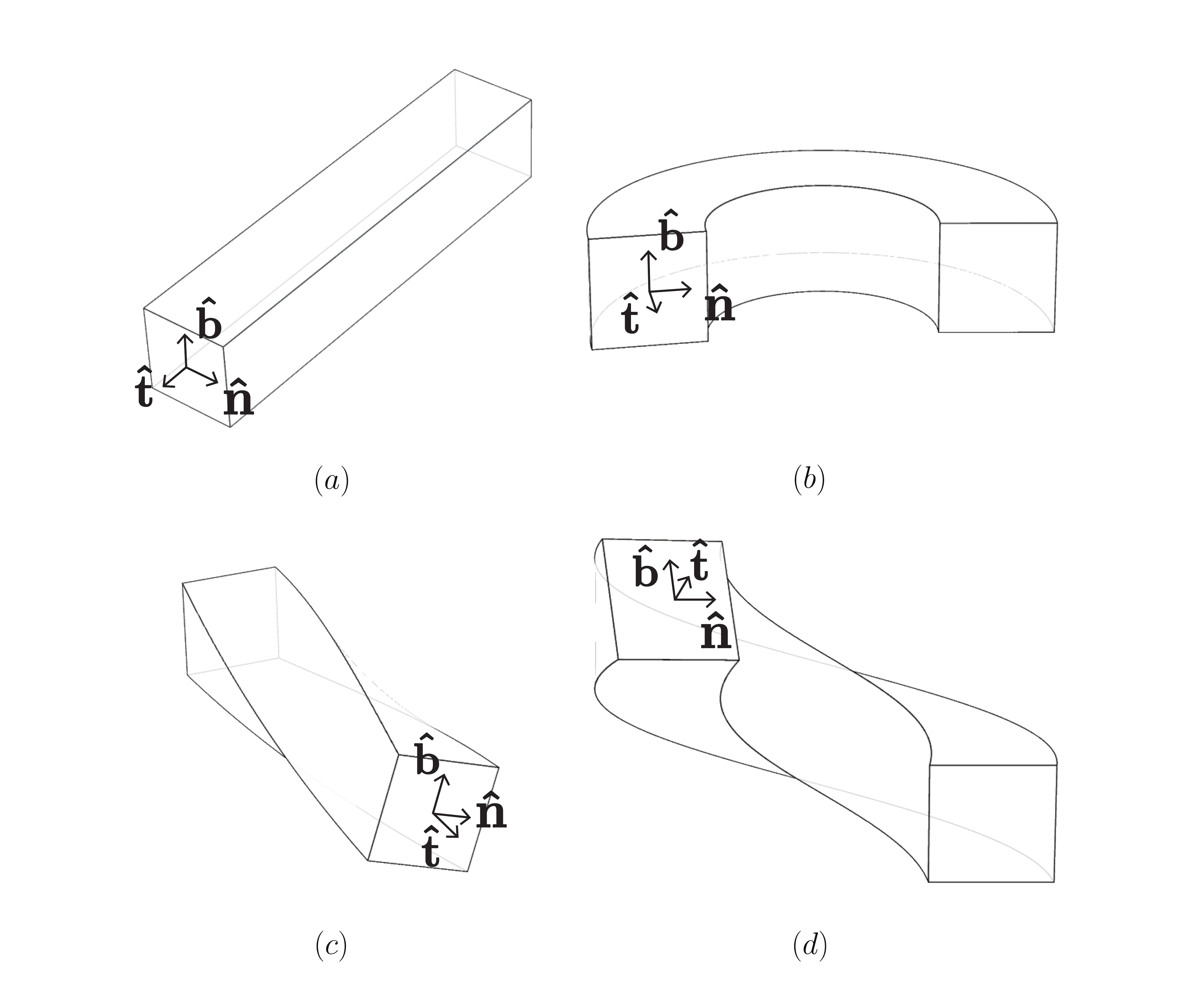}
\caption{The  geometries with corresponding curvature $\kappa$ and torsion $\tau$. (a) Straight duct with $\tau = \kappa = 0\,\text{mm}^{-1}$, (b) toroidal duct with $\tau = 0 \,\text{mm}^{-1}$ and $\kappa = 1/3\,\text{mm}^{-1}$, (c) twisted duct with $\tau = 1/8 \,\text{mm}^{-1}$ and $ \kappa = 0\,\text{mm}^{-1}$, and (d) helical duct with $\kappa = 1/3\,\text{mm}^{-1}$ and $\tau = 1/8 \,\text{mm}^{-1}$. 
} 
\label{fig:models} 
\end{figure*}

To characterize the contributions of curvature and torsion to the flow field, we simulated fluid flow in four different geometries (Fig. \ref{fig:models}).  
We defined the geometries using established methods \citep{Bolinder1996,Bolinder1996a}, which are summarized here.
The centerlines of the geometries are described by the position vector $\mathbf{r} = \mathbf{r}(s)$ and parametrized with the arc length $s$.
To facilitate the decomposition into axial and transverse flow components later on, we use the orthonormal Frenet-Serret frame consisting of the tangent ($\mathbf{\hat{t}}$), normal ($\mathbf{\hat{n}}$), and binormal ($\mathbf{\hat{b}}$) unit vectors:
\begin{align}
\mathbf{\hat{t}}  =  \mathbf{r'}, \quad 
\mathbf{\hat{n}}  = \frac{\mathbf{\hat{t}'}}{\lVert \mathbf{\hat{t}'} \rVert},   \quad  \textrm{and} \quad  
\mathbf{\hat{b}}  = \mathbf{\hat{t}}  \times \mathbf{\hat{n}},
\end{align} 
where the prime indicates the derivative with respect to the arc length $s$.
The curvature $\kappa$ and torsion $\tau$ of the centerline are defined as 
\begin{align}
\kappa  = \left\lVert \mathbf{\hat{t}'} \right\rVert \quad \textrm{and} \quad  
\tau = \mathbf{\hat{n}'} \cdot \mathbf{\hat{b}}.
\end{align}  
For the helical centerline, we obtain
\begin{align}
\kappa = \dfrac{R}{K^2+R^2} \quad \textrm{and} \quad
\tau = \dfrac{K}{K^2+R^2} \quad,
\end{align}
where $R$ is the radius and $2 \pi K$ is the pitch of the helix, i.e., the height of a complete helical turn. 
The toroidal duct (Fig. \ref{fig:models}b) has a centerline with zero torsion and a curvature of $1/R$.
In contrast, the centerline of the twisted duct (Fig. \ref{fig:models}c) has zero curvature and a torsion of $1/K$.
For our geometries, the centerline curvature and torsion were chosen as $\kappa = 1/3\,\text{mm}^{-1}$ and $\tau = 1/8\,$mm$^{-1}$, respectively, to match average properties observable in human cochleae \cite{Viergever1978,Wimmer2019human}.
To save computational cost, we reduced the total arc length of the centerline for all models to $s_{max}= 10\,$mm, which is shorter than a human cochlea ($37\,$mm) \citep{rask2012human}. 
The models have a $2\,$mm$\,\times\,2\,$mm cross-section to capture typical dimensions of the cochlear cross-section \citep{aebischer2021}.

\subsection{Model Implementation}
The perilymph flow in the cochlea is modelled as flow of an incompressible Newtonian fluid \citep{Steele2014}, by the Navier-Stokes equations
\begin{align}
\begin{split}
    \nabla \cdot  \mathbf{u} &= 0 \quad \textrm{and}\\
    \rho \frac{\partial \mathbf{u}}{\partial t} + \rho (\mathbf{u} \cdot \nabla) \mathbf{u}  &= - \nabla p + \mu \Delta\mathbf{u}, 
    \label{eq:NS}
\end{split}
\end{align}
where $\mathbf{u} = \mathbf{u}(x,y,z,t)$ and $p = p(x,y,z,t)$ are the velocity and pressure fields at time $t$ represented in an Eulerian reference frame.
The dynamic viscosity and the fluid density are denoted by $\mu$ and $\rho$, respectively (Table \ref{tab:properties}).
At the inlet and outlet surfaces, the pressure $p_{\rm{out}}$ and $p_{\rm{in}}$ were set to
\begin{align}
\begin{split}
p_{\rm{in}}(t) &= P_0 \cos( 2 \pi f t )    \quad  \textrm{and} \quad  
p_{\rm{out}}(t) = 0,
\label{eq:boundaries}
\end{split}
\end{align}
with the oscillation frequency $f$ and the pressure amplitude $P_0$.
No-slip boundary conditions were imposed on the walls. 
We performed simulations at 12 frequencies ranging from 0.125$\,$Hz to 256$\,$Hz in powers of two.
This frequency range corresponds to infrasound ($<$$\,16\,$Hz) and the human low-frequency hearing regime (16$\,$Hz to 256$\,$Hz) and covers quasi-steady (0.125$\,$Hz) to unsteady inertial flows for which the associated Womersley number is greater than unity (Table \ref{tab:properties}). 
We chose this frequency range for the following reasons. 
First, the low stimulation frequencies enable us to verify our results in quasi-steady-state scenarios with the known solutions of \citet{Bolinder1996}. 
Second, low frequencies are perceived in the apical region of the cochlea, which exhibits the strongest geometric curvature and torsion. 
Third, higher frequencies ($>256\,$Hz) require refined meshes due to steeper velocity gradients that would make the calculations prohibitively expensive. 

\begin{table}[ht]
\centering
\caption{Model parameters. The viscosity and density are taken from water at 37$^\circ$C (body temperature). The inlet pressure amplitude $P_0$ was chosen such that a desired mean velocity amplitude $W_0 $ (averaged over the cross-section) was obtained.}
\vspace{2mm}
\begin{adjustbox}{width=0.5\textwidth}
\begin{tabular}{lr}
\toprule
\textbf{Parameter}                      & \textbf{Value} \\
\midrule
Dynamic viscosity $\mu$  	            &  $0.69\,$\si{mPa.s}  \cite{Steele2014}\\  
Density $\rho$                          & $993\,$\si[per-mode=symbol]{\kilogram\per m^3} \cite{Steele2014} \\
Hydraulic diameter $d_h$                & $2\,$mm \citep{aebischer2021}\\
Mean velocity amplitude  $W_0 $   & $2-200\,$\si[per-mode=symbol]{\micro\meter\per\second} \citep{Greene2017,aibara2001human,koch2022methods} \\
Reynolds number $Re = d_h W_0  \rho/\mu$  & \qquad $0.0058   -  0.58 $\\
Oscillation frequency $f$ \qquad        & $0.125 - 256\,$Hz\\
Womersley number $\alpha = \frac{d_h}{2} \sqrt{ 2 \pi f \rho / \mu} $   \qquad  &  $1 - 48$\\
\bottomrule
\end{tabular}
\end{adjustbox}
\label{tab:properties}
\end{table}

\subsection{Numerical Model}

\begin{figure}[htbp!]
\centering
{\includegraphics[width=0.4\textwidth]{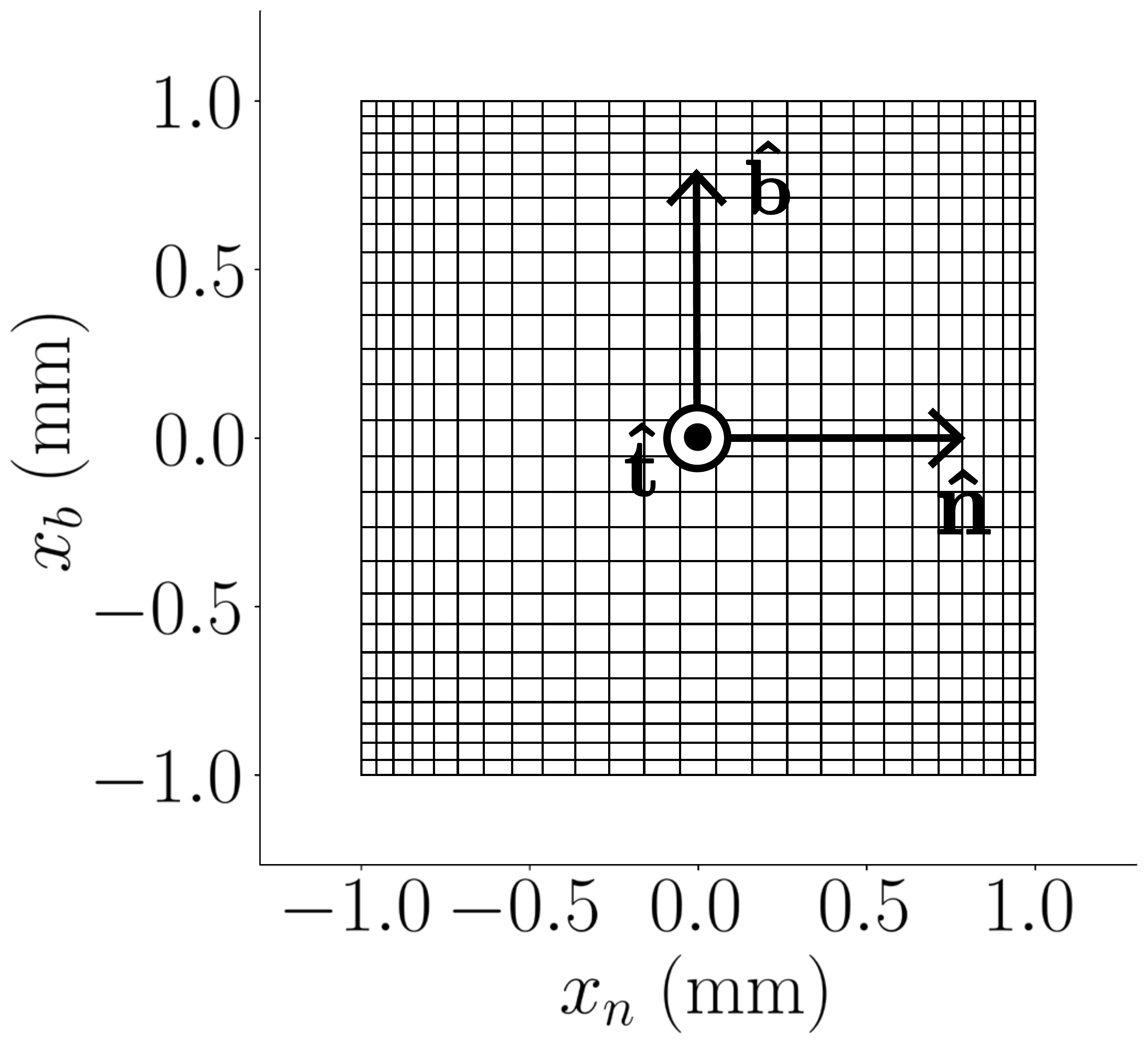}}\qquad\quad\hfill
\caption{Cross-sectional mesh with $26 \times 26$ nodes along the normal  $\mathbf{\hat{n}}$ and binormal $\mathbf{\hat{b}}$ direction (Frenet-Serret frame).} 
\label{fig:grid_rec} 
\end{figure}

For the simulations, we used the finite element solver COMSOL Multi\-physics\textsuperscript{\tiny\textregistered} (COMSOL AB, Stockholm, Sweden). 
Structured meshes were generated by sweeping a square cross-section (Fig. \ref{fig:grid_rec}) along the ducts' centerlines.
The cross-sectional mesh was aligned along the centerline with $\mathbf{\hat{t}}$, $\mathbf{\hat{n}}$, and $\mathbf{\hat{b}}$ (see section \ref{ch:DuctGeometries}). Grid stretching was applied.
The number of nodes was chosen based on a convergence analysis, resulting in $26 \times 26$ nodes in the cross-section and a total of $77\,500$ hexahedral elements per geometry.
We used Lagrange elements of order two and one for the velocity and pressure (P2P1), respectively.
For the time-dependent solver, the implicit backward differentiation formula method of variable order (between 1 and 5) was used. 
We chose 100 steps per oscillation period of the pressure boundary condition.
The inlet pressure amplitude $P_0$ was found iteratively such that the axial velocity amplitude averaged over the cross-section $W_0$ remained the same across the stimulation frequency range for the different geometries.

Based on reported data \citep{Greene2017,aibara2001human,koch2022methods}, we chose three mean inlet velocity amplitudes averaged over the cross-section: \SI[per-mode=symbol]{2}{\micro\meter\per\second}, \SI[per-mode=symbol]{20}{\micro\meter\per\second}, and \SI[per-mode=symbol]{200}{\micro\meter\per\second}.
For a constant velocity amplitude $W_0$, decreasing frequencies correspond to increasing equivalent sound pressure levels at the ear drum. 
The characteristic middle ear transfer function (i.e., the stapes velocity versus the sound pressure at the ear drum) reaches its resonance at about $1\,$kHz, with a mean slope of approximately 6$\,$dB per octave up to 1$\,$kHz \citep{aibara2001human,koch2022methods}.
For example, at 32$\,$Hz, a stapes velocity of \SI[per-mode=symbol]{200}{\micro\meter\per\second} can be expected for an external auditory canal pressure of $\sim 101\,$dB sound pressure level (SPL).
At 128$\,$Hz, the same stapes velocity would correspond to an acoustic stimulus with $\sim 113\,$dB SPL. 
Velocity amplitudes of \SI[per-mode=symbol]{20}{\micro\meter\per\second} and \SI[per-mode=symbol]{2}{\micro\meter\per\second} match pressure levels that are 20$\,$dB and 40$\,$dB lower, respectively.
The selected stapes velocities approximately cover the range of sound pressure levels occurring during normal conversations (\SI[per-mode=symbol]{2}{\micro\meter\per\second}), shouted conversations (\SI[per-mode=symbol]{20}{\micro\meter\per\second}), and near to the threshold of pain at 256$\,$Hz (\SI[per-mode=symbol]{200}{\micro\meter\per\second}).
The selected parameters result in Reynolds numbers $Re$ well below unity, implying that the fluid phenomena are in the Stokes regime. 

To shorten the duration of the initial transient state, the amplitude of the inlet pressure $P_0$ was ramped up smoothly over the first few cycles.
%(see Appendix \ref{A:SimulationTime}). 
To ensure that the initial transient is washed out, we retrieved results after a sufficient number of cycles, e.g., after 21 cycles at 256$\,$Hz.
%(Table \ref{tab:simutime}). 
The computations were performed on a computing cluster with AMD Epyc2 processors.  
The typical turnaround times for the computations that were running with 16 tasks and 32$\,$GB RAM per CPU on one node varied from one day for 0.125$\,\text{Hz}$, to seven days for 256$\,\text{Hz}$.
\subsection{Axial and Transverse Flow}
\label{sec:PrimarySecondaryFlow}
%To avoid ambiguity, we use the terms of axial flow and transverse flow to characterize velocity field components.
We apply the Frenet-Serret frame to decompose the velocity field $\mathbf{u}$:
\begin{equation}
\mathbf{u} = u \mathbf{\hat{n}} + v \mathbf{\hat{b}} + w \mathbf{\hat{t}},
\label{eq:expansion}
\end{equation}
and define the velocity component along the tangent as axial flow $w = \mathbf{u} \cdot \mathbf{\hat{t}}$, while the components in the normal and binormal directions, $u = \mathbf{u} \cdot \mathbf{\hat{n}}$ and $v = \mathbf{u} \cdot \mathbf{\hat{b}}$, constitute the transverse flow.
 %Note, because the frame is orthonormal, the transverse flow is always perpendicular to the axial flow.

\section{Results}
\subsection{Axial Flow}
\begin{figure*}[hbt!]
\hspace*{-0.5cm}
\centering
{\includegraphics[height=0.92\textheight]{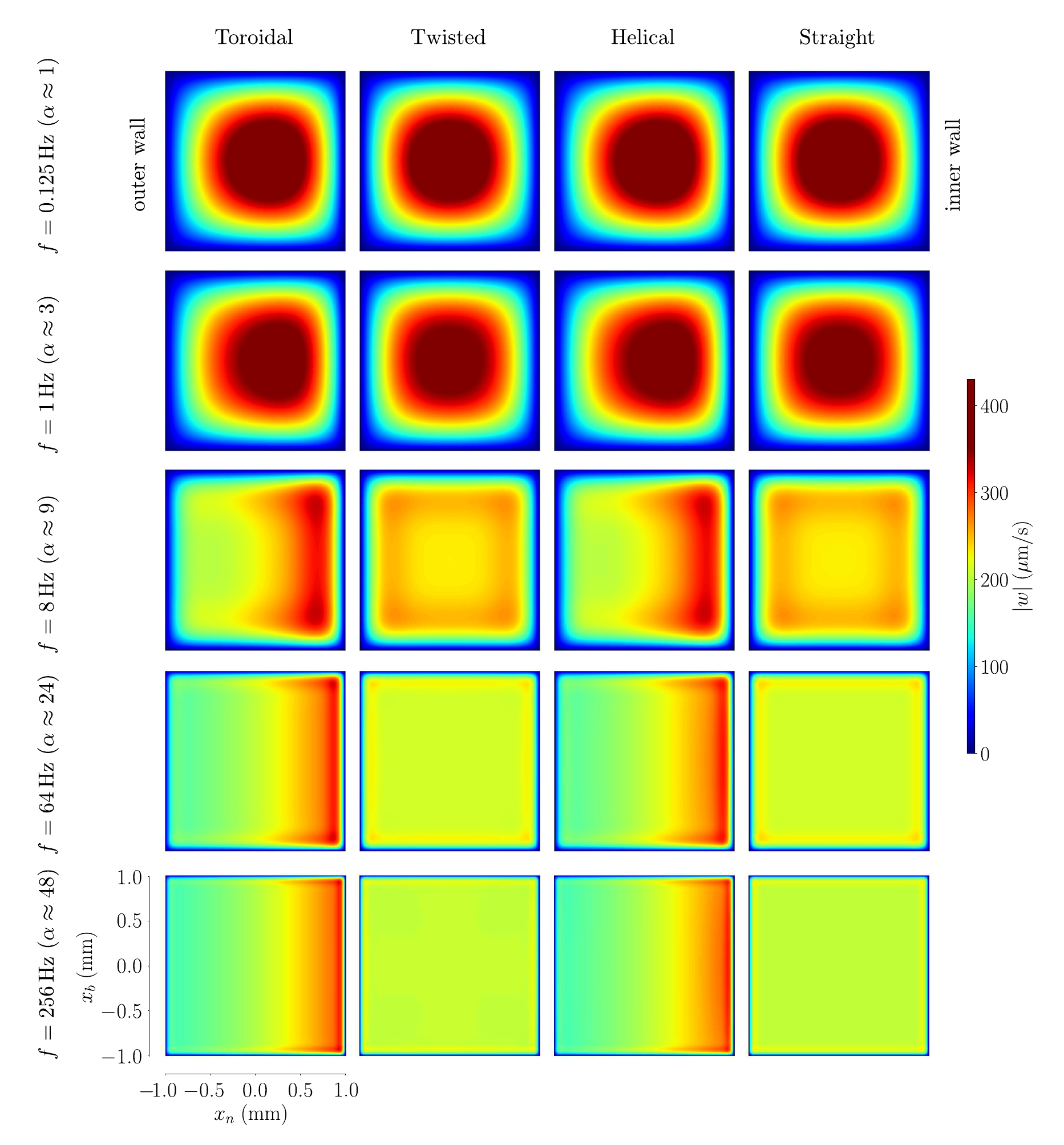}} 
\caption{Axial flow magnitude for different frequencies and geometries shown at the peak amplitude ($W_0 = 200\,\si[per-mode=symbol]{\micro\meter\per\second}$).}
\label{fig:VelProfileall}
\end{figure*}

Figure \ref{fig:VelProfileall} shows the axial velocity contours in the geometries for different oscillation frequencies taken at the peak amplitude.
For the straight duct, the results are in good agreement with the exact solution by \citet{Tsangaris2003}. 
The relative velocity error with respect to their solution (averaged in the cross-section) is 0.09\% at 0.125$\,$Hz and increases to 0.25\% at 64$\,$Hz and 1.25\% at 256$\,$Hz.
As expected for Womersley flow, the axial velocity profiles develop thin boundary layers and steep gradients at the walls with increasing Womersley numbers \citep{Womersley1955}.

\begin{figure}[hbt!]
\centering
{\includegraphics[width=0.95\textwidth]{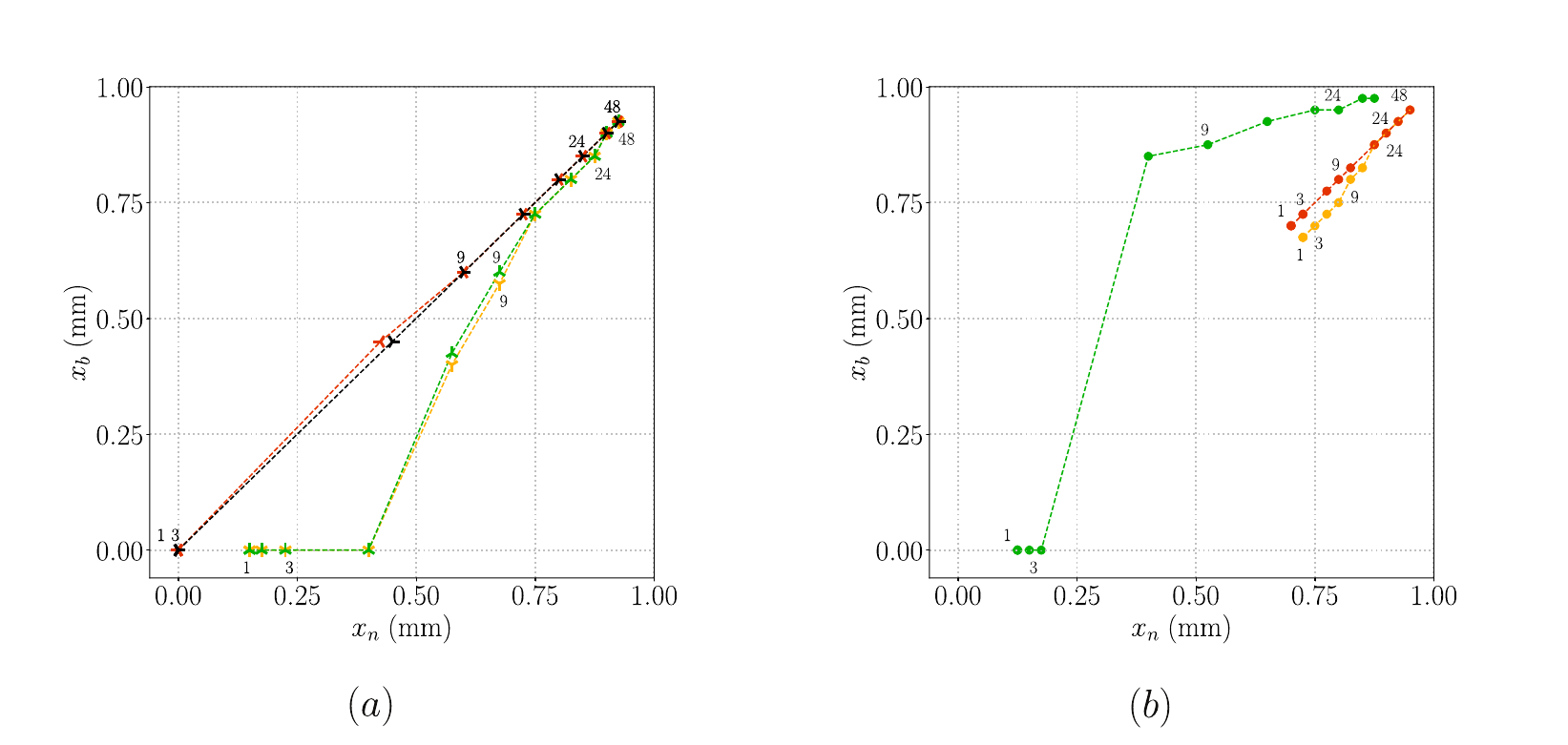}}
\caption{Location of maximum axial (a) and transverse flow (b) in the toroidal (green), twisted (red), helical (yellow)  and straight (black) ducts for Womersley numbers ranging from 1 to 48 and a mean velocity amplitude of $W_0=200\,\si[per-mode=symbol]{\micro\meter\per\second}$. Only the first quadrant of the cross-section is shown.}
\label{fig:positionMax}
\end{figure}

The axial velocity in the twisted duct hardly differs from the one in the straight duct (relative mean difference of $0.39$\%).
Also, the flow profile in the helical ducts differ only little from the toroidal duct profiles ($0.51$\%).
Figure \ref{fig:positionMax}a shows the locations of the maximum axial flow in the cross-section for different frequencies and geometries. 
With increasing Womersley numbers $\alpha$, these locations converge toward the corners in the straight and the twisted ducts. 
The maximum axial velocities in the toroidal and helical geometry are shifted to the inner wall corners for low Reynolds numbers, as also observed by \citet{Pantokratoras2016} and \citet{Murata1976}. 

In the straight and twisted ducts, the axial velocity profile has a central symmetry with respect to the normal and binormal directions. 
The axial velocity profiles in toroidal ducts are symmetrical about the horizontal axis at $x_b=0$, while in the helical duct a small deviation is introduced by the non-zero pitch. 

\FloatBarrier
\subsection{Transverse Flow}
\begin{figure*}[hbt!]
\hspace*{0.5cm}
\centering
{\includegraphics[height=0.92\textheight]{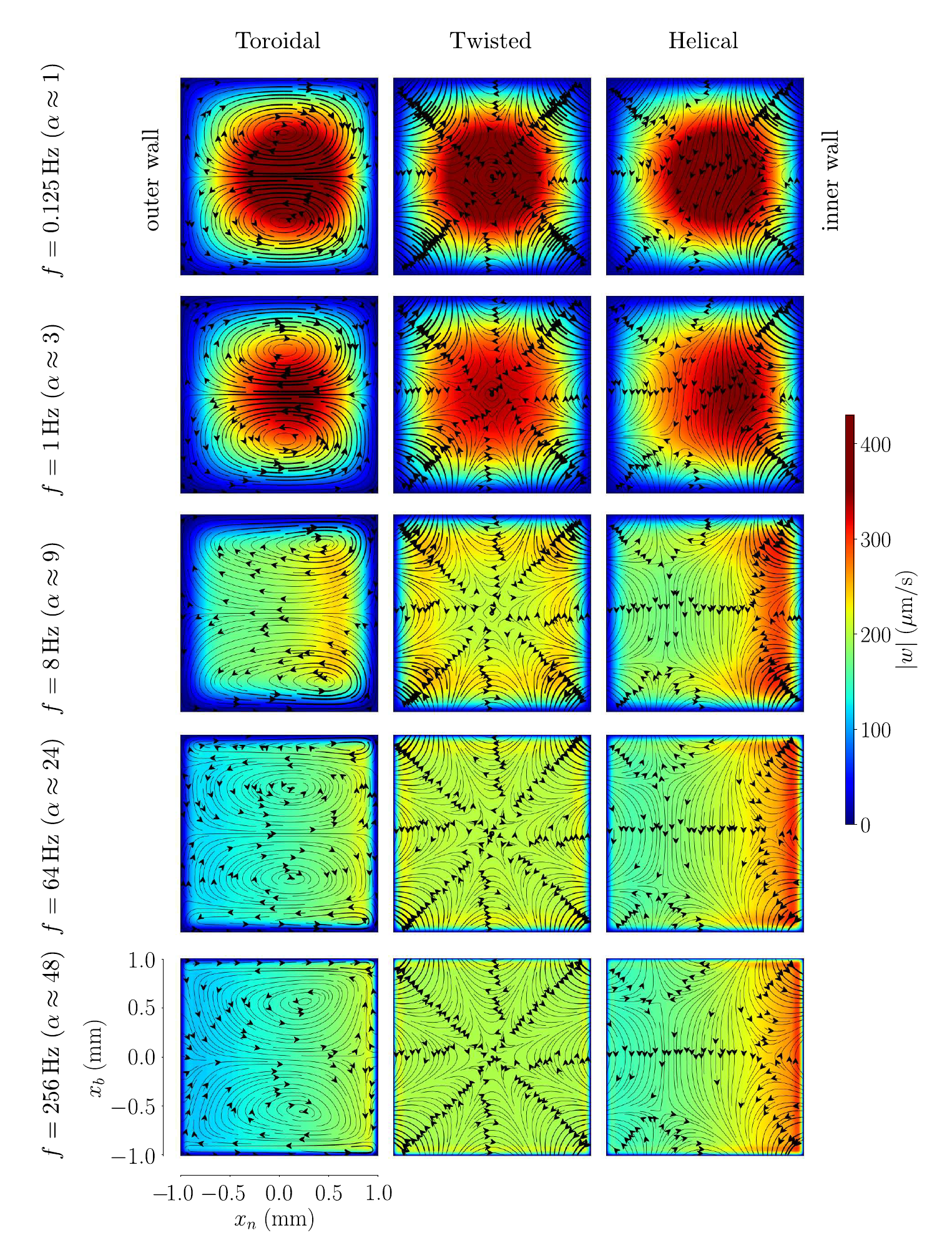}} 
\caption{Axial (colored contours) and transverse flow (streamlines) for different frequencies and geometries shown at the time of maximum transverse flow ($W_0 = 200\,\si[per-mode=symbol]{\micro\meter\per\second}$). }
\label{fig:streamplots}
\end{figure*}

Figure \ref{fig:streamplots} shows transverse flow patterns for selected oscillation frequencies (additional figures  are provided in the Supplemental Material at [URL will be inserted by publisher]).
In the quasi-steady situation (i.e., 0.125$\,$Hz), we found similar patterns as described by \citet{Bolinder1996}, who studied transverse flow in steady conditions.

In the toroidal duct, two counter-rotating Dean cells form, caused by inertial forces that push the fluid along the middle line ($x_b=0$) from the inner towards the outer wall \citep{Dean1928}. 
This effect leads to unidirectional transverse flow, regardless of the direction of the axial flow. 
As the oscillation frequency increases, the Dean cells separate and concentrate on the walls.
At 32$\,$Hz ($\alpha \approx 17$) and higher frequencies, an additional pair of vortices, called Lyne instabilities, can be observed in the center of the channel, rotating opposite to the Dean cells \citep{lyne1971unsteady}.

The torsion of the twisted duct causes saddle flow in the corners of the cross-section, alternating with the direction of the axial flow.
The observed pattern stays relatively stable over the simulated frequency range.
The transverse flow pattern in the helical duct shows a similar structure as in the twisted duct, except that the transverse flow is focussed toward the inner wall,  where the axial flow is stronger as well. 

\begin{figure*}[hbt!]
\centering
\includegraphics[width=1\textwidth]{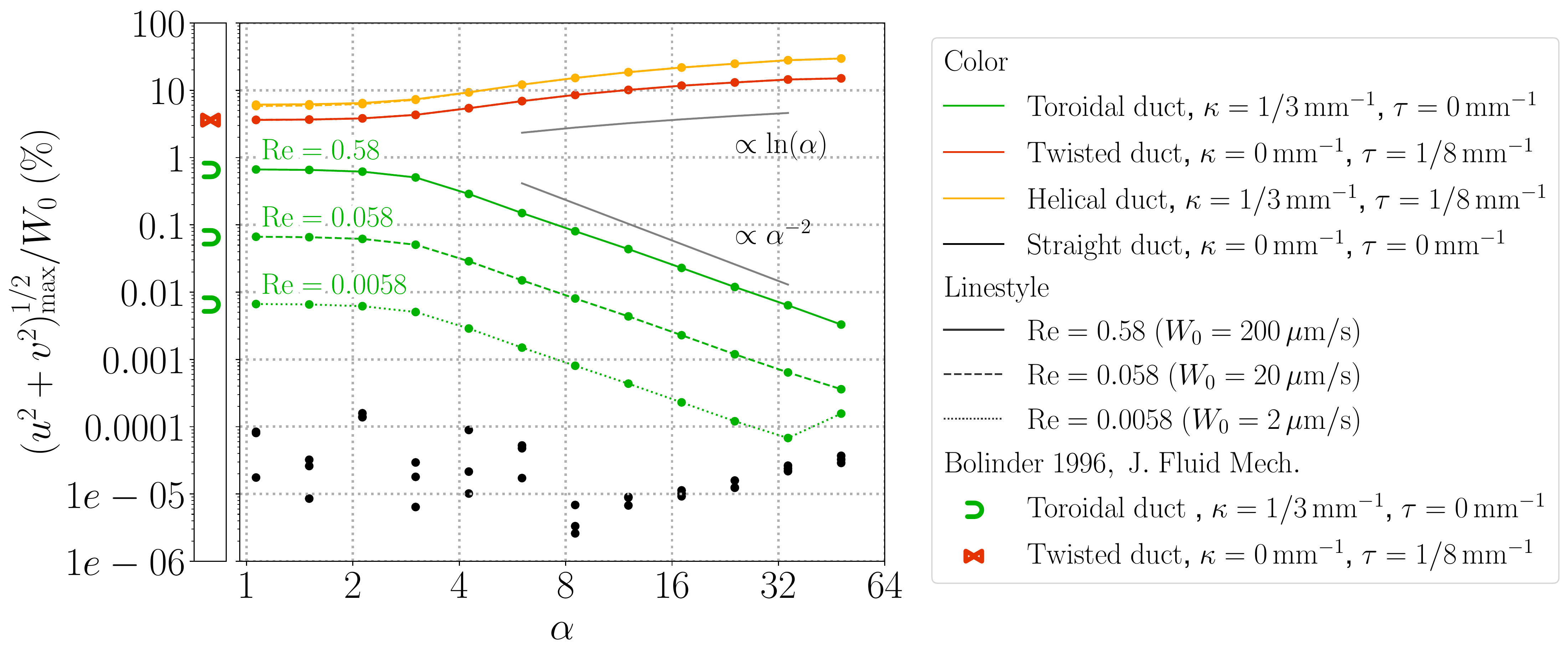}
\caption{Maximum transverse flow magnitude with respect to the mean axial flow velocity ($W_0$) as a function of the Womersley number $\alpha$. 
The markers on the left side of the figure show the steady flow solutions for the toroidal and twisted ducts \citep{Bolinder1996}.
Black symbols indicate numerical noise defined as the maximum transverse flow velocities found in the straight duct simulations. Note that the maximum transverse flow at $\alpha \approx 48$ ($f=256 \,$Hz) and $Re = 0.0058$ in the toroidal duct is distorted due to its proximity to the numerical noise floor.}
\label{fig:SecFlowWo}
\end{figure*}

In general, for all geometries, the location of the maximum transverse flow starts close to the center of the cross-section, but approaches the corners (of the inner wall, in helical and toroidal ducts) for increasing Womersley numbers (Fig. \ref{fig:positionMax}b).
Figure \ref{fig:SecFlowWo} summarizes the  maximum transverse velocity magnitude versus Womersley number in our geometries.
As expected for the toroidal duct, the transverse flow magnitudes because of curvature are small compared to the axial velocity (below $1$\% of the axial velocity magnitude).
This is because the magnitude of the transverse flow in a toroidal duct is proportional to  $Re^2$ \citep{Bolinder1996}. 
Increasing the oscillation frequency further reduces the relative magnitude of the transverse flow which scales roughly with $f^{-1}$ (or  $\alpha^{-2}$).
As the Womersley number increases, the contribution of curvature to the transverse flow continues to decrease and approaches negligible amplitudes in the toroidal duct.

As pointed out by \citet{Bolinder1996} for steady flow at low Reynolds numbers, torsion has a dominating effect on transverse flow, which can be seen by a 5-fold higher maximum transverse flow magnitude in the quasi-steady state (i.e., 0.125$\,$Hz).
Strikingly, in contrast to curvature, the contribution of torsion is substantial and gets more significant with increasing frequencies following approximately $\textrm{ln}(\alpha)$, with the relative maximum exceeding 10\% of the axial flow magnitude for Womersley numbers greater than 10. 
At 256$\,$Hz ($\alpha \approx 48$), the maximum transverse flow velocity accounts for approximately 15\% of the axial flow velocity.

For the twisted duct, the curves for the different Reynolds numbers are coinciding in Fig. \ref{fig:SecFlowWo}. 
This agrees with \citet{Bolinder1996} who found that the magnitude of transverse flow scales to first order with $Re$ in such geometries. 
Also for the helical duct the curves for different $Re$ coincide, indicating that the maximum transverse flow remains proportional to $Re$, at least in the frequency range observed in our experiments. 

Surprisingly, although the contribution of curvature alone is negligible in the toroidal duct, the combination of curvature with torsion results in a 2-fold gain of transverse flow magnitude in the helical duct.
This gain can be observed throughout the simulated frequency range, even for cases with $\alpha > 10$, where the isolated contribution of curvature to the transverse flow is less than $0.1$\% of the axial flow magnitude.

\subsection{Phase Lag}
\begin{figure*}[hbt!]
\centering
\includegraphics[width=1\textwidth]{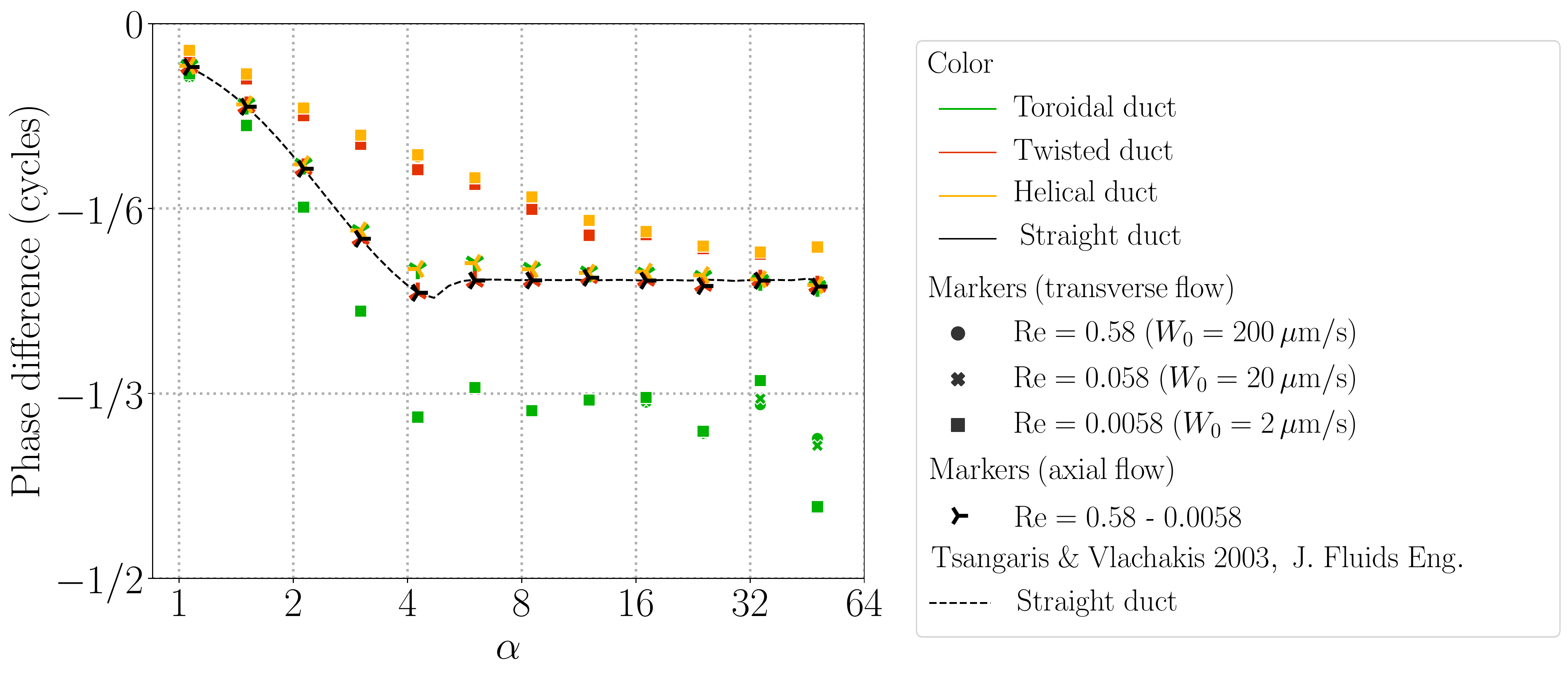}
\caption{Phase difference between maximum axial (trilateral markers) and transverse flow velocity (marker style according to the Reynolds number) and pressure for different Womersley numbers $\alpha$. The black dashed line indicates the exact solution (phase lag between the axial flow and pressure) for a straight square duct \citep{Tsangaris2003}. Markers for different Reynolds numbers $Re$ coincide. For the toroidal duct, the phase difference is distorted at $\alpha \approx 48$ ($f=256 \,$Hz) and $Re = 0.0058$ due to the numerical noise floor.}
\label{fig:phaselagmaxsec_grid} 
\end{figure*}

Figure \ref{fig:phaselagmaxsec_grid} shows the phase difference between the flow and pressure oscillations. 
As the phase depends on the location within the cross-section, the reported values are shown at the maximum amplitudes for the axial (Fig. \ref{fig:positionMax}a) and transverse flow (Fig. \ref{fig:positionMax}b).
The axial velocity phase lag in the straight duct is in good agreement with the exact solution by \citet{Tsangaris2003}.
The axial phase lag in the twisted duct resembles the one in straight ducts (relative difference of $0.15$\%), reaching a plateau at a quarter of a cycle for $\alpha > 4$. 
In the toroidal and helical ducts, the phase lag is similar but on average higher than in the straight and twisted ducts.
 
The phase lag of the transverse flow velocity decreases monotonically in all geometries and approaches the phase lag of the axial flow velocity with increasing Womersley numbers, except for the toroidal duct. 
In the toroidal duct the phase lag fluctuates around a third of a cycle for $\alpha > 4$. 
%At 16$\,$Hz, this corresponds to a time difference of approximately $20\,$ms, which decreases to $1.3\,$ms at 256$\,$Hz.

\section{Discussion}
\subsection{Transverse Flow}
In this study, we used simplified geometries to investigate the effects of curvature and torsion on transverse flow under oscillatory stimulation based on conditions in the human cochlea.
Our models, which have square cross-sections and do not include fluid interactions with flexible membranes, serve as abstract representations of the cochlear scalae. 
Through our simulations, we discovered two key findings about the influence of geometry on transverse flow. 

First, we found that torsion remains a significant contributor to transverse flow in unsteady conditions at low Reynolds numbers ($<1$), as pointed out by \citet{Bolinder1996} for steady flow, and its contribution increases with higher oscillation frequencies. 
The formation of saddle flow due to torsion could be interpreted as a kinematic effect, as the twisted geometry forces a change of direction close to the boundaries.

Second, we observed that the combined effect of curvature and torsion on transverse flow is greater than what would be expected based on a simple combination of the two effects, especially when the relative contribution of curvature alone becomes negligible.
The transverse velocities resulting from curvature are about one order of magnitude smaller than found from torsion, and decrease rapidly for increasing frequencies. 
The transverse flow patterns are similar between the twisted and the helical ducts (Fig. \ref{fig:streamplots}), however, caused by curvature, the maximum axial velocity gets shifted toward the corners of the inner duct wall in helical ducts (Fig. \ref{fig:positionMax}b), leading to stronger saddle flow there. 
Therefore, it appears as if the helical shape acts as an amplifier to transverse flow when stimulated with oscillations within the observed frequency range.

\subsection{Axial Flow}
The straight and twisted ducts have nearly identical axial flow profiles in our simulations.
This is in agreement with \citet{Kheshgi1992} who found that the axial flow profile in twisted channels approximates the velocity profile of a straight channel under steady conditions and low torsion.
As soon curvature is present (i.e., in toroidal and helical ducts), the axial flow profiles are dominated by curvature over the observed frequencies.
At low Reynolds numbers, the maximum velocity is shifted towards the inner wall of the bend, contrary to what one would expect for higher Reynolds numbers \citep{Pantokratoras2016,Murata1976}.  
With increasing Womersley numbers, the maximum flow velocity moves closer to the inner wall.
In our geometries, torsion of the helical duct (or the resulting pitch) has little effect on the axial flow compared to the toroidal duct. 

\subsection{Phase Lag}
We found that the phase difference depends on the duct geometry for transverse flow and behaves differently than the phase difference between th pressure and axial flow velocity. 
In toroidal ducts, the axial and transverse flow velocities can be considered in-phase when approaching higher frequencies, with a quarter of a cycle offset to the pressure amplitude.
The difference in the axial phase lag between the geometries with a straight centerline and the ones with a curved can be explained by the different probing locations, which were closer to the walls for helical and toroidal ducts (Fig. \ref{fig:positionMax}a).
Also the fluctuations in the transverse phase lag in the toroidal duct for $\alpha>4$ likely result from the change in the probing locations with increasing Womersley numbers  (Fig. \ref{fig:positionMax}b).
 
\subsection{Study Limitations}
The main limitation of our study is the use of abstract geometries to model the highly complex anatomy of the human cochlea \citep{DePaolis2017}.
Further studies on models featuring more realistic cross-sections and tapering, and in particular fluid-membrane interactions, are required to investigate possible implications on cochlear mechanics.
Our simulations are limited to the observed frequency range, which only covers the low-frequency hearing regime of humans. 
Additional simulations, with refined mesh geometries, are required to obtain data covering the whole human hearing range up to 16$\,$kHz and more.
 
\section{Acknowledgements}
We would like to thank Prof. Nathaniel Greene, University of Colorado, for providing middle ear transfer function data. 
We acknowledge the support from the Swiss National Science Foundation (Grant No. 205321\_200850).
Calculations were performed on UBELIX (www.id.unibe.ch/hpc), the HPC cluster at the University of Bern.

%\bibliography{references} 
%
%
\providecommand{\noopsort}[1]{}\providecommand{\singleletter}[1]{#1}%
\end{document}